\documentclass{emulateapj}
\usepackage{apjfonts}

\def\hi{\ifmmode {\rm H}\,{\sc i}~ \else H\,{\sc i}~\fi}

\def\zphot {z_{\rm phot}}

\slugcomment{Published as  ApJ 713, (2010), 738-750}

\shorttitle{Star formation and the galaxy mass-size relation}
\shortauthors{Williams et al.}

\begin{document}

\title{The evolving relations between size, mass, surface density, and star
formation in $3\times 10^4$ galaxies since $z=2$}

\author{Rik J. Williams\altaffilmark{1,2}, 
	Ryan F. Quadri\altaffilmark{1}, 
	Marijn Franx\altaffilmark{1}, 
	Pieter van Dokkum\altaffilmark{3},
	Sune Toft \altaffilmark{4,5}, 
	Mariska Kriek\altaffilmark{6}, 
	Ivo Labb\'e\altaffilmark{2,7}} 
\altaffiltext{1}{Leiden Observatory, Leiden University, Niels Bohrweg 2, 
NL-2333 CA  Leiden, The Netherlands}
\altaffiltext{2}{Carnegie Observatories, Pasadena, CA 91101, USA}
\altaffiltext{3}{Department of Astronomy, Yale University,
                 New Haven, CT 06520-8101, USA}
\altaffiltext{4}{Dark Cosmology Centre, Niels Bohr Institute, University
of Copenhagen, Juliane Maries Vei 30, DK-2100 Copenhagen, Denmark}
\altaffiltext{5}{European Southern Observatory, Karl-Schwarzschild-Str. 2,
D-85748 Garching bei M\"unchen, Germany}
\altaffiltext{6}{Department of Astrophysical Sciences, Princeton University,
Princeton, NJ 08544, USA}
\altaffiltext{7}{Hubble Fellow}

\email{williams@obs.carnegiescience.edu}

\begin{abstract}
The presence of massive, compact, quiescent galaxies at $z>2$ 
presents a major challenge for theoretical models of galaxy formation
and evolution.  Using one of the deepest large public near-IR surveys
to date, we investigate in detail the correlations between star
formation and galaxy structural parameters (size, stellar mass, and 
surface density) from $z=2$ to the present.
At all redshifts, massive quiescent galaxies (i.e. those with little or no
star formation) occupy the extreme high end of the surface density 
distribution and follow a tight mass-size correlation, while star-forming 
galaxies show a broad range of both densities and sizes.  Conversely,
galaxies with the highest
surface densities comprise a nearly-homogeneous population with little
or no ongoing star formation, while less dense galaxies exhibit
high star-formation rates and varying levels of dust obscuration.
Both the sizes and surface densities of quiescent galaxies 
evolve strongly from $z=2-0$; we parameterize this evolution for
both populations with
simple power law functions and present best-fit parameters for comparison to
future theoretical models.  Higher-mass quiescent galaxies undergo faster
structural evolution, consistent with previous results.  
Interestingly, star-forming galaxies' sizes
and densities evolve at rates similar to those of quiescent galaxies.
It is therefore 
possible that the same physical processes drive the structural evolution of 
both populations, suggesting that ``dry mergers'' may not be the sole culprit
in this size evolution.
\end{abstract}



\section{Introduction}
Ample evidence now exists for the presence of massive
galaxies at $z>2$ with little or no ongoing star formation, suggesting
that a non-negligible fraction of the local early-type galaxy population
was effectively in place only a few Gyr after the big bang, and that 
massive red galaxies constitute a significant fraction of (or possibly
even dominate) the stellar mass density at $z\sim 2$
\citep[e.g.][]{mccarthy04,labbe05,daddi05,kriek06,kriek08a,rudnick06,marchesini07,stutz08,toft09}.  
Although their masses and star formation rates (SFRs) are nominally similar, 
it has since become clear that these quiescent galaxies
at $z>1$ have dramatically different structures compared to local 
ellipticals: specifically, the high-redshift galaxies' sizes are much
smaller 
\citep[e.g.][]{daddi05,trujillo06a,zirm07,toft07,vdokkum08,cimatti08,damjanov09}.  Their
surface densities likewise evolve strongly with redshift, as does the
``threshold'' surface density above which galaxies are predominantly
quiescent \citep{franx08,maier09}.  These phenomena pose several
new challenges for studies of galaxy formation.  Indeed,
if massive quiescent galaxies at $z\sim 2$ were almost universally
compact and dense, why are similar objects in the local
universe practically nonexistent \citep[e.g.][]{trujillo09,taylor09b}?

Several mechanisms for this transformation have been proposed, including
a scenario whereby local
early-type galaxies are built through a series of ``dry'' (i.e., gas-poor) 
mergers of these compact progenitors \citep{khochfar08,hopkins08,feldmann09}.  Such 
mergers, being largely
non-dissipative, would over time tend to ``puff up'' compact galaxies.
Some observational estimates of the major dry merger rate since $z\sim 1$
indicate that these are indeed a factor in the buildup
of massive galaxies \citep[][]{bell06,bundy09,deravel09},
but the remnants of such major mergers would also have correspondingly
larger masses and so it is unclear whether major dry mergers can entirely 
solve the size-mass discrepancy 
\citep[though they are likely to mitigate it to some degree; e.g.][]{vdwel09}.  
Minor mergers may be a more important process -- both from a 
simple virial argument \citep{naab09} and because
massive compact ``cores'' at high redshift may accrete material in 
their outskirts to form the massive ellipticals seen today
\citep{bezanson09,hopkins09,vdokkum10}.   However, theoretical models have
yet to converge on a definitive explanation -- some proposed models 
predict galaxy sizes dramatically different from those observed
\citep{joung09}, while others match observed sizes at $z=2$ and $z=0$ 
\citep{khochfar06} but not at $z\sim 1$ \citep[]{vdwel08}.  

Whatever the underlying mechanism may be,
observations of large samples of galaxies over a broad redshift 
range are crucial
to adequately test current and upcoming models.  However, many of the
aforementioned galaxy structure studies employ HST-NICMOS, which 
provides accurate size measurements but can typically image only one target 
at a time and is thus inadequate for observing very large samples.  
Optical imagers are much larger, but at $z\sim 2$ near-IR
data are more robust for size determinations; these bands fall in 
the rest-frame optical and therefore better trace the distribution
of stellar mass than the rest-frame UV (observed optical).
Furthermore, spectroscopy of quiescent galaxies (both to determine their
redshifts and confirm their quiescence) requires large amounts of
observing time due to the absence of strong emission lines; again, 
while this can be performed for a few bright galaxies at $z\sim 2$
with current facilities \citep{kriek06}, it is infeasible
for the large samples of fainter objects needed to provide a comprehensive
view of the high-redshift galaxy population.

Fortunately, although spectroscopy and space-based imaging give the
most precise picture of star formation and galactic structure for
individual objects, the average properties of large samples can
be accurately investigated with less ``expensive'' data.
For example, both \citet{trujillo06b} and \citet{franx08} successfully
based their galaxy size measurements on ground-based imaging and
photometric redshifts.
Although there are inherently greater uncertainties on the size and mass of 
any individual object than with NICMOS imaging and spectroscopic 
redshifts, the {\it average} structural 
parameters of large populations can be accurately determined.  
Furthermore, even without spectroscopy there are several 
ways to determine which galaxies are quiescent and which 
are actively forming stars.  One straightforward method is to
identify galaxies on the ``red sequence'' (using, e.g., the rest-frame
$U-V$ color), which at low redshift is primarily composed of quiescent 
galaxies with strong 4000\AA\ breaks.  However, star-forming
galaxies containing large amounts of dust can have similarly red colors,
so ``red and dead'' galaxy samples selected through this color cut are
likely to be contaminated with substantial numbers of ``red and dusty'' 
starbursts with increasing redshift \citep[e.g.][]{williams09}.

With photometric observations in a suitable set of filters, the shape of the 
broadband spectral
energy distribution (SED) can be used to distinguish between the sharp
4000\AA\ break characteristic of old stellar populations and the more gradual
reddening caused by dust absorption; the best-fit SFRs
and dust column densities from SED modeling codes can then be used
to define quiescent galaxy samples.  Empirical 
selection techniques using multiple rest-frame colors, such as the one 
employed by \citet{williams09}, are also effective at
separating dusty from ``dead'' galaxies up to $z=2$ (and even higher; 
I.~Labb\'e et al., in preparation) with little dependence on input templates or 
models.  Coupled with the large, deep multiwavelength surveys and improved
photometric redshift techniques that have emerged in the past few years,
these diagnostics provide an unprecedented view of star formation and
its relation to other physical properties in massive galaxy populations.

Using the largest such publicly-available data set with sufficient
depth, the UKIDSS Ultra-Deep Survey (and overlapping
imaging from the Subaru-XMM Deep Survey and Spitzer Wide-Area
Extragalactic Survey), we analyze the structural evolution of quiescent 
galaxies up to $z=2$, specifically the correlations between star formation
activity, galaxy size, and stellar mass surface density.
First in \S\ref{sec_data} we review the data and describe the size and 
mass measurements
employed for this study.  Next in \S\ref{sec_ssfr} we investigate how
the specific star-formation rate (sSFR) is related to galaxy mass, size, and
surface density, and in \S\ref{sec_rfsize} present
quantitative constraints on the evolution of these structural properties 
for both star-forming and quiescent galaxies.  
Finally, the uniformity of the connections between star formation and
structure is discussed in \S\ref{sec_uniformity} along with some novel
ways to interpret these correlations.
AB magnitudes and a concordance cosmology ($h=0.7$, $\Omega_M=0.3$, 
$\Omega_\Lambda=0.7$) are assumed throughout.

\section{Data and Measurements} \label{sec_data}
The high-redshift galaxy sample analyzed herein is based on an updated
version of the 
$K$-selected galaxy catalog\footnote{Available from\\
\url{http://www.strw.leidenuniv.nl/galaxyevolution/UDS}} described in detail 
by \citet{williams09}; a brief summary follows.  
This catalog comprises near-infrared $J$ and $K$ data taken from
the UKIDSS Ultra-Deep Survey Data Release 1 
\citep[UDS;][]{lawrence07,warren07},
with overlapping $BRi^\prime z^\prime$ imaging from the Subaru-XMM
Deep Survey \citep[SXDS;][]{sekiguchi04}, and 
$3.6/4.5\mu$m data from the Spitzer Wide-area Infrared Extragalactic
(SWIRE) survey \citep{lonsdale03}.  Using the positions and shapes of
bright but unsaturated stars in the different bands, all mosaics were
astrometrically matched and convolved to consistent point-spread functions (PSFs).
Using the SExtractor software \citep{bertin96}, ``color'' fluxes were
measured in fixed 1\farcs 75 apertures and total $K$ fluxes from flexible
elliptical \citep{kron80} apertures.

In addition to the original \citet{williams09} catalog, we incorporate the 
$H$-band mosaic from the UKIDSS Data Release 3 (to be described by S.~J.~Warren 
et al., in preparation) with PSF- and astrometric matching and flux 
measurements performed in the same manner as with the other bands.  
Photometric redshifts were calculated with the $H$ data included
using the publicly-available code EAZY \citep{brammer08}.  The new
photometric redshifts differed very little from the original ones
presented by \citep{williams09}, but inclusion of the $H$
band should in principle improve both the $\zphot$ values and stellar
mass estimates at $z\ga 2$.  At the adopted flux limit of $K<22.4$,
this catalog contains nearly $3\times 10^4$ galaxies.

Because of its depth and $\sim 0.8$\,deg$^2$ area, the UDS is well-suited for
studies of large galaxy samples at $z\ga 0.5$; at lower redshifts, however, 
the comoving volume probed by the UDS is too small to provide meaningful
samples (and may also be severely affected by cosmic variance).  Thus,
for comparison to the UDS we also include
the same $z\sim 0.06$ SDSS sample used by \citet{franx08}.  To summarize,
this sample was originally defined by \citet{kauffmann03}, and \citet{franx08}
performed minor corrections in the mass-to-light ratios (accounting
for the fact that the galaxy centers, where the spectroscopic fibers
are placed, are typically redder than the outskirts) as well as total
flux corrections.  The redshift range of $z=0.05-0.07$ was chosen to 
avoid selection effects on galaxies with high mass (at the low-$z$ end) and 
small angular sizes (at higher redshift).

\subsection{Sizes} 
\subsubsection{Fitting}
Size measurements were performed in a manner similar to that
by \citet{toft07}.  We use the Galfit software package \citep{peng02} 
to fit \citet{sersic68} models to all bright ($K<22.4$) sources 
detected in the unconvolved (i.e., before PSF matching, to ensure the highest
possible angular resolution) UDS $K$ image.  This image had a typical
seeing FWHM of 0\farcs7.  First a square postage stamp 
21 pixels (4\farcs 2) on a side was made around each galaxy to be fitted.  Initial
guesses for the effective radius $r_e$, magnitude, ellipticity, and position 
angle were taken from the SExtractor catalog.  The PSF used by Galfit to
deconvolve the galaxy images is the average PSF over the image (taken
as the median of 300 bright, unsaturated stars; also used in the PSF-matching
step described in Williams et al.~2009).  The S\'ersic parameter $n$ was
allowed to vary between 1 and 4 and the effective radius between 0\farcs01
(effectively a point source) and 15\arcsec.  Since galaxies in general may
exhibit color gradients, it is important to measure sizes at approximately 
the same rest-frame wavelength when comparing samples at different
redshifts.  The procedure described above was thus repeated for the 
unconvolved $J$ and $H$ images (both exhibiting 0\farcs8 seeing), and we 
define the size of a galaxy as its 
circularized effective radius ($r_e=\sqrt{ab}$) at a rest-frame
wavelength of 8000\AA, interpolated from the size measurements of the
two adjacent bands.  In other words, when rest-frame 8000\AA\ falls
between two observed bands $i$ and $j$, the interpolated size is
\begin{equation}
r_{e,8000}=r_{e,i}+\frac{r_{e,j}-r_{e,i}}{\lambda_j-\lambda_i}\left[8000\rm{\AA}(1+z)-\lambda_i\right]
\end{equation}

In practice the interpolated sizes are quite 
similar to those computed by, e.g., simply taking the size measured from 
the observed band closest to 8000\AA.  However, the interpolation method
smoothes out possible discretization effects due to the choice of the measurement
filter, and also in principle helps to mitigate failed size measurements
in a given band.  

\begin{figure}
\plotone{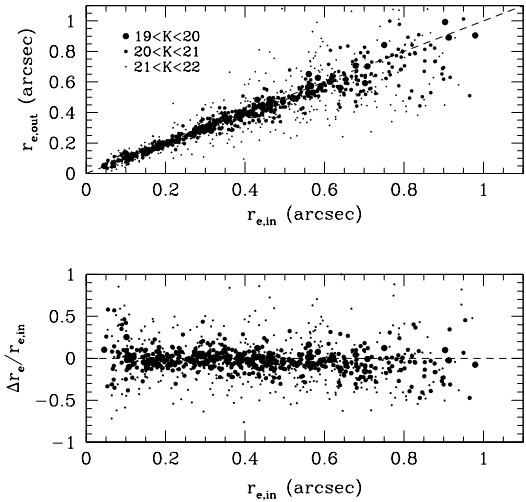}
\caption{{\it Top panel:} Output effective radii of simulated galaxies as a
function of $r_{e,in}$.  Larger point sizes represent brighter galaxies;
for clarity only a random 10\% of our simulated galaxies in these magnitude
bins are plotted.
{\it Bottom panel:} Fractional deviation 
$(r_{e,out}-r_{e,in})/r_{e,in}$ as a function of input radius.
Fainter galaxies exhibit larger random uncertainties.  Although the
faintest, largest galaxies show a small systematic offset ($10-20$\% at 
worst), these deviations are small compared to the random scatter and the
measured trends with redshift discussed in \S\ref{sec_rfsize}.
\label{fig_sims}}
\end{figure}

\subsubsection{Simulations}
Previous studies have shown that reliable galaxy sizes can be 
measured from ground-based data, provided the signal-to-noise ratio is 
sufficiently high and the PSF across the image is stable 
\citep[e.g.][]{trujillo06b,franx08}.  For instance, 
\citet{trujillo06b} use simulations to investigate
systematic effects, finding them to be minimal down to 
$r_e\sim 0.2-0.3\times$\,FWHM(PSF), though individual measurements exhibit
a large degree of scatter.  The effective radius is
typically the most robust structural parameter measured; other 
variables defining the profile shape (e.g.~Sersi\'c index and axis ratio) 
are far more susceptible to systematic uncertainties.

It is nonetheless instructive to investigate whether 
the specific characteristics of the UDS data and our
fitting techniques have introduced systematic biases;
we thus performed simulations analogous to \citet{trujillo06b} 
to test our effective radius measurements.  Model galaxy profiles over
a range of magnitudes $K=18-23$, $r_e$ from 0.1 to 1 arcsec, and $n=1-4$ were 
created; these were chosen to span typical galaxy angular sizes at 
$z=1-2$.  Blank postage stamps (containing
only noise) were then cut from random positions in the UDS $K$-band image,
and the model galaxies added to them.  The identical fitting procedure
used on the real images was then applied to 15000 such
simulated postage stamps to derive the output effective radii.  

Figure~\ref{fig_sims} shows the distribution of input 
and output effective radii, and the fractional difference between the
two, as a function of input effective radius.  Brighter galaxies are
plotted as larger points.  As expected, relatively faint galaxies exhibit
larger random uncertainties in their size measurements.
While systematic deviations are seen for both the faintest and largest
galaxies, these are relatively small (with a median of $\sim 10$\%) 
compared to the
random errors; additionally, since we only include galaxies with
$z>0.5$ in our analysis, objects with such large angular sizes are quite
rare.  These observed offsets are comparable to the offsets reported
by \citet{trujillo06b}; most importantly, the average effective
radii of galaxies with sizes $\sim 0\farcs 1-1\farcs 0$ appear
to be reasonably reliably recovered.  Given the large scatter, however,
it should be noted that individual measurements of galaxy sizes 
are highly uncertain even if ensemble averages are accurate.

\subsubsection{Empirical consistency tests} 
The simulations described above provide a realistic view of some of the
random and systematic uncertainties in this analysis.  However, to some
extent these are idealized; for example, the evident ability to measure
effective radii as small as one-half of a pixel may be true in a numerical
sense, but in real images additional effects such as PSF variations will
limit the precision that can be achieved.
We therefore supplement the simulations with a set of purely empirical
tests to better assess the reliability of the galaxy size measurements.
Specifically, we check the following three important assumptions:
(1) the simple interpolation of galaxy sizes between different bands
is robust; (2) a single PSF is sufficient to model the entire image; 
and (3) the fitting box
size of 4\farcs2 is large enough to obtain reasonable fits.  

First, to test the interpolation assumption, we estimated the 
$H$-band sizes by averaging the measured $J$ and $K$ sizes; the 
estimated and measured 
$H$ sizes matched very well with no noticeable systematic offset.  The
``single PSF'' assumption (2) was tested by generating four different PSFs, 
each composed
of bright stars from the four quadrants of the $K$ mosaic, and re-measuring
the sizes of galaxies within the central $40\arcsec\times 40\arcsec$ of
the $K$ mosaic.  While there
were small systematic differences in the sizes measured with the different
PSFs, these were on the order of 0\farcs02, which is effectively negligible
compared to typical measured galaxy sizes of 0\farcs2-1\farcs0.  Similarly,
the median galaxy sizes show no evidence of systematic deviations
with position in the image, indicating that the PSF does not significantly 
vary.  Finally, the box-size assumption (3) was tested by re-fitting a
subset of galaxies  using postage stamp sizes of 5\farcs2 and
6\farcs2; the new sizes were fully consistent with those measured in the 
original 4\farcs2 cutouts.  The exception to this is the largest objects 
in this sample with $r_e\ga 1\farcs 2$ (about 10\,kpc radius at $z=1-2$); 
these objects exhibit ~30\%\ larger sizes when a larger fitting box is 
used, but such objects comprise only a tiny fraction ($<4$\%) of our 
$z>0.5$ sample and so do not affect our results.

Re-fitting the same subset of galaxies using different PSFs and box sizes
also gives an estimate of the measurement uncertainties (i.e., the degree
to which a given size measurement is reproducible given different 
assumed input parameters).
Between 0\farcs2 and 1\farcs0 the measured sizes were consistent with each
other within $\sim 5$\%; at 0\farcs1 the fractional error rapidly
increases to about 10\%$-15$\%, and at smaller effective radii sizes can no 
longer be reliably inferred.  To determine the efficiency to which Galfit can
distinguish between point sources and extended objects in these data,
we also measured the radii of stars  (found via color selection) using 
the same fitting parameters. Interestingly, while essentially all of the
stars have best-fit radii of zero (as is expected for point sources), 
Galfit measures non-zero radii for the vast majority of compact galaxies -- 
even those with best-fit radii much smaller than the PSF.

Thus, these simulations and empirical tests confirm that effective radius 
measurements are likely free of major systematic effects down to 
$r_e= 0\farcs1-0\farcs2$.  Even if some offsets are
present \citep[as seen in our simulations and by][]{trujillo06b}, 
they are comparable to or smaller than the systematic uncertainties 
on other parameters (e.g., in mass, redshift, and SFR
determinations).  We therefore conclude that
measurements of the average sizes of large galaxy samples down to small radii
are robust given these data.  This is similar to what \citet{franx08} 
found by fitting their ground-based data and comparing it to Advanced
Camera for Surveys (ACS) imaging.

\subsection{Stellar masses and star-formation rates}
The stellar masses of galaxies in this sample were calculated with the 
{\it Fitting and Assessment of Synthetic Templates} (FAST) code
\citep{kriek09}.  This code uses $\chi^2$ minimization
to fit \citet{bc03} stellar population evolution models to the observed 
broadband photometry.  A \citet{salpeter} initial mass function (IMF) 
and solar metallicity were 
assumed; masses were then shifted by a factor of $-0.2$\,dex for 
consistency with the $z\sim 0$ Sloan Digital Sky Survey (SDSS) masses (which
were computed with a \citet{kroupa01} IMF).  We
re-fitted a subset with \citet{maraston05} models and both IMFs 
to verify that this factor is essentially constant and does not 
vary systematically with galaxy color or mass.  
Redshifts were fixed to the $\zphot$ values derived by
EAZY, and a variety of evolutionary histories
were allowed in the fitting, including exponentially-declining
(SFR$\sim e^{-t/\tau}$) models with $\tau$ ranging from $10^{7-10}$\,Gyr.
The computed masses are consistent with those calculated using other
standard methods, for example using HYPERZ as a fitting 
engine and scaling the model amplitudes to estimate masses
\citep[as done by, e.g.,][]{forster04}.  Note that \citet{maraston05}
models result in galaxy masses approximately $0.2$\,dex \emph{lower} than 
\citet{bc03} models; however, unlike the IMF correction factor, this 
discrepancy varies somewhat with galaxy mass.

Other parameters, such as the degree of dust extinction $A_V$, 
average stellar population ages, exponential factor $\tau$, and
SFR are also computed by FAST during the SED fitting
procedure.  Most of these are highly uncertain when based on
broadband data \citep{kriek08b}; however, the 
{\it sSFR} (sSFR$=$SFR$/M_\star$) is somewhat more
robust.  This quantity,
taking the mass of the host galaxy into account, provides a more
meaningful characterization of star-formation activity (e.g., an SFR
of 10$\,M_\odot$/year is far more significant in a dwarf galaxy
than a giant elliptical).  Furthermore, since the total SFR
and mass both exhibit a similar dependence on the assumed IMF, 
the sSFR is less sensitive to the choice of IMF and input
stellar population model(s) than the absolute SFR.

\begin{figure}
\plotone{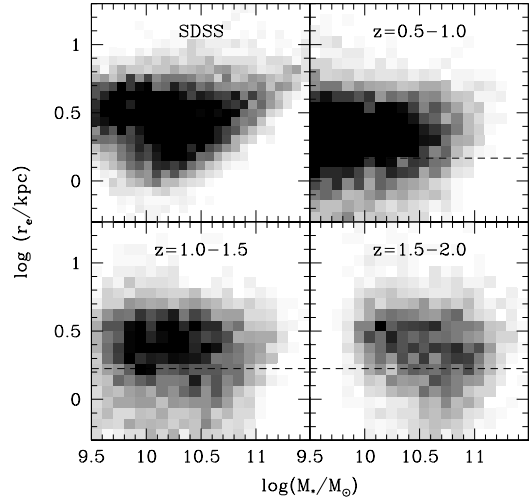}
\caption{Size-mass relation for galaxies at $z\sim 0.06$ (SDSS; {\it
upper left}) and three redshift ranges in the UKIDSS UDS.  The
grayscale denotes the number of points within each bin; dashed lines 
indicate the angular radii ($\theta\sim 0\farcs 2$) below which
individual size measurements have larger uncertainties.  Massive, 
compact galaxies
(i.e.~those in the lower right-hand portion of each panel) are almost
nonexistent at $z=0$ but become progressively more common at higher
redshifts.  The radii of galaxies with $M>10^{10.8}$\,M$_\odot$ 
evolve roughly as $r_e\sim (1+z)^{-0.89}$ (see also Table~\ref{tab_fit}).
\label{fig_smassbw}}
\end{figure}

\begin{figure*}
\epsscale{1.15}
\plotone{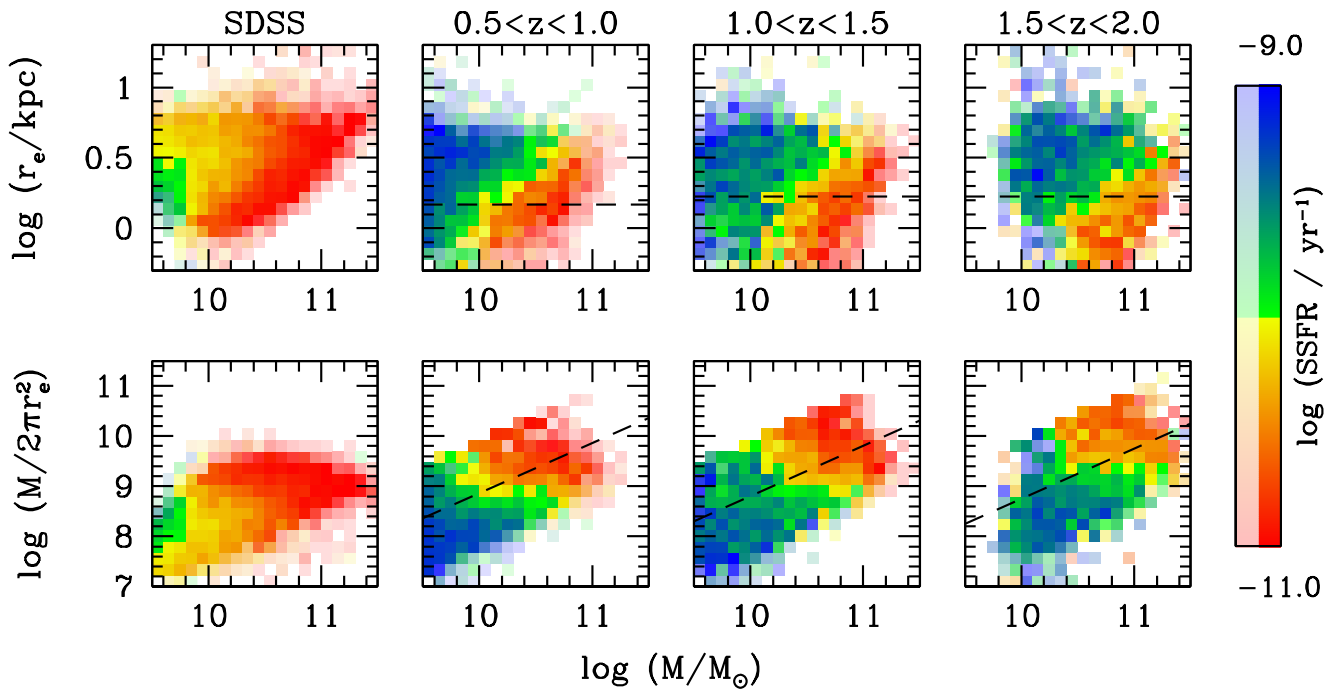}
\caption{{\it Top:} Size-mass relation in SDSS (left panel), 
and in three UDS redshift bins: $z=0.5-1$, $z=1-1.5$, and $z=1.5-2$ in the
second, third, and fourth panels respectively.  
{\it Bottom:} Surface density-mass relation in the same redshift bins.
In all plots, the color denotes the mean sSFR
of the underlying galaxies; ``faded'' bins
contain two or fewer galaxies.  Dashed lines indicate apparent effective radii
of 0\farcs2, below which (or above the corresponding surface
density) individual galaxy size measurements have relatively large 
uncertainties (though the binned averages are more reliable).  Strong 
evolution is evident in the mass-size and 
mass-surface density relations of low-sSFR galaxies, and the compact, dense
galaxies which are clearly present in the early universe gradually vanish
with decreasing redshift.  
\label{fig_smass}}
\end{figure*}

\subsection{Rest-frame fluxes}
Rest-frame colors, particularly the combination of 
$U-V$ and $V-J$, are invaluable for distinguishing quiescent galaxies from 
those actively forming stars \citep{wuyts07,williams09}.  Although such
colors can be derived directly from the templates used for photometric
redshift fitting, by definition these colors are confined to the range of 
colors spanned by the template set (which may be narrower than the intrinsic
range of the galaxy population).  Instead, we interpolate 
rest-frame $U$, $V$, and $J$ fluxes directly from the observed photometry
and photometric redshifts using the same method as 
\citet{williams09} but with the new $H$-band data included. This is
accomplished using the InterRest utility \citep{taylor09a}, which in
turn is an implementation of the method described by \citet{rudnick03}.
The reddest observed band in our data set is IRAC $4.5\mu$m, 
corresponding to rest-frame $J$ at $z\sim 2.5$; reasonably robust
fluxes can thus be interpolated up to approximately this redshift.

\subsection{Completeness}
In general, the mass completeness limit of a flux-limited galaxy 
sample depends strongly on color: since red galaxies have higher 
mass-to-light ratios, mass-limited red galaxy samples become incomplete 
at brighter magnitudes than for blue galaxies.  Completeness limits 
derived for strongly heterogeneous (mixed red and blue) samples are therefore 
dominated by blue galaxies, and such samples can still exhibit serious 
incompleteness effects among massive red galaxies.  Computing mass
limits from red galaxies alone therefore provides a more conservative
estimate of the overall sample completeness.

We derive mass completeness limits for the UDS sample by comparing the masses
of red galaxies ($U-V_{\rm rest}>1.5$) to their observed $K$ 
magnitudes.  The 75\% completeness limit is then defined as the mass at
which about 25\% of the galaxies in the sample fall below the adopted
$K$ flux limit.  This is performed for the $1.5<\zphot<2.0$ redshift
bin, thus ensuring that our analysis does not suffer from significant
incompleteness effects up to $z=2$.  To ensure
accurate size measurements, the size and surface density analysis
is restricted to galaxies with $K_{\rm lim}<22.4$ \citep[approximately 
1 magnitude brighter than the formal $5\sigma$ survey limit, also used
in][]{williams09}, corresponding to a mass limit of $\log M_\star>10.6$.

\section{The relation between mass, size and star formation} \label{sec_ssfr}

\subsection{Mass-size relation} 

Our newly constructed sample allows us to study the mass-size relation
and its evolution to $z=2$.  
This relation, in three redshift bins from the UDS, is plotted
in Figure~\ref{fig_smassbw}.  To supplement the higher-redshift
data, the same relation from SDSS at $z\sim 0$ is also shown.

Two notable trends stand out: first, there is a correlation
between radius and mass, whereby the most massive galaxies out to $z=1.5$
have on average larger sizes than less massive galaxies; this effect
is weaker or nonexistent at $z=1.5-2$.  A second and related point is
that there are effectively no massive, compact galaxies 
($M>10^{11}$\,M$_\odot$,
$r_e\la 2$\,kpc; in the lower right-hand region of the size-mass plot) 
at $z=0$, but at higher redshifts this area
of parameter space becomes progressively more populated.  This simply
reflects the strong size evolution of massive galaxies found in several
previous studies and introduced in \S1.

\subsection{Star formation as a function of mass and size}
\citet{kauffmann03,kauffmann06}  and \citet{franx08} showed that
the broad range in galaxy effective radii, at a given mass, is tightly 
correlated with sSFR out to $z=2.5$.
These authors concluded that sSFR is a tight
function of stellar mass surface density ($M_\star/R^2$), or velocity dispersion
\citep[$\sqrt{M_\star/r_e}$;][]{franx08}.
Put differently, these results imply that the size-mass relation is
different for galaxies with different sSFRs.

Our sample is ideally suited to study this aspect at higher
redshifts, as it covers an area $\sim 18$ times
larger than that used by \cite{franx08}.
We show in Figure~\ref{fig_smass} how the sSFR
depends on mass and size.
It is clear that the dependence is very strong: at a given mass, galaxies with
low sSFRs (binned and plotted as red/yellow squares
in this figure) have small sizes, while those
with large sizes have high sSFRs (green/blue squares).
The effect is not only strong at very low redshift, but extends to the
highest redshift bins.
At the same time, the sizes of galaxies with low sSFRs
(red in the figure) are very small at high redshift, consistent
with results obtained by others on smaller samples
\citet{toft07,vdokkum08,franx08,toft09}.

Table~\ref{tab_smass} lists the best-fit slopes and normalizations
of the \emph{quiescent galaxy} size-mass relations seen in 
Figure~\ref{fig_smass}.
For this we defined ``quiescent'' as galaxies exhibiting 
SSFR$<0.3/t_H$, where $t_H$ is the age of the universe at each galaxy's
redshift, effectively picking out the red and yellow points in 
Figure~\ref{fig_smass} (see also \S\ref{sec_rfsize} and the Appendix).  In each redshift
interval the median log effective radius was calculated in mass bins of
width 0.2\,dex, only including galaxies above the mass completeness limit 
at each redshift, and a least-squares fit was performed to these median
points.  For convenience, the normalizations of the power-law fits 
in Table~\ref{tab_smass} are defined as the log effective radius of 
a typical galaxy with $M=10^{11}M_\odot$.  Uncertainties were computed
using bootstrap resampling.  Note that any comparisons
between the mass-size relations seen in SDSS and UDS should
be treated with caution, as systematic differences in the size and mass
determinations between these two samples may exist.  Nonetheless, strong 
evolution in the
normalization of the size-mass relation is evident; the slope of the
relation, however, does not significantly change between $z=0.5-2$.

As the strong mass-size correlation of quiescent galaxies implies,
sSFR does not depend
strictly on size or mass alone, but rather on some combination of
the two.
The bottom panels of Figure~\ref{fig_smass} instead show how sSFRs depend
on stellar mass surface density ($\Sigma_\star=M/2\pi r_e^2$) and mass.  
Galaxies with the  highest surface densities have weak star formation, 
with very little dependence on mass.  In addition, we can define a
``threshold surface density'' between quiescent and star-forming galaxies
\citep[similar to that defined by][]{franx08} at each redshift; in
Figure~\ref{fig_smass}, the sharp color division
(between green and yellow) at log\,sSFR$=-10.0$ could be taken as a simple 
example of one such threshold.  This clearly increases with increasing
redshift.  Note that this particular threshold depends only weakly on mass
(i.e. is nearly horizontal in the $\Sigma$-mass plot), confirming that
surface density is more fundamentally related to star formation activity
in galaxies than mass \citep{franx08}.

Furthermore, this figure clearly shows
at which epoch galaxies of a given size, mass, and density were forming stars: 
for example, galaxies with $M=10^{11}$\,M$_\odot$ were generally star-forming
at $z\sim 2$ but by $z=1.0-1.5$ most of their star formation activity
had ceased.  However, the most compact massive galaxies (the
lower right-hand region of the radius-mass plot) are quiescent
at all redshifts considered here, and must have been quenched at $z>2$.

Most intriguingly, the compact quiescent galaxies visible at high redshift
do not maintain their small sizes or high surface densities to lower
redshifts; instead, the quiescent galaxies progressively
evolve to larger radii and lower surface densities with decreasing
redshift.  This phenomenon is apparent from the UDS data alone in
Figure~\ref{fig_smass}, but
is most starkly illustrated by the $z\sim 0$ SDSS panel where essentially
no massive, compact galaxies are visible.  This re-illustrates the
phenomenon pointed out by \citet{vdokkum08}, \citet{trujillo09}, and
\citet{taylor09b}, but
with multiple redshift bins the depletion of these galaxies is now
easily seen.  Since these galaxies are unlikely to be losing more than
a small fraction of their mass \citep[via mass loss from evolved 
stars;][]{damjanov09}, size evolution must be the primary
driver behind their disappearance.

\begin{deluxetable}{lcc}
\tablecolumns{3}
\tablewidth{200pt}
\tablecaption{Fits to the quiescent galaxy size-mass correlations shown
in Figure~\ref{fig_smass} \label{tab_smass} }
\tablehead{
\colhead{Redshift} &
\colhead{A} &
\colhead{b}
}
\startdata
SDSS  & $0.634\pm 0.004$ &$0.41\pm 0.01$\\
$0.5<z<1.0$    &$0.46\pm 0.02$ &$0.54\pm 0.06$\\
$1.0<z<1.5$    &$0.35\pm 0.01$ &$0.56\pm 0.06$\\
$1.5<z<2.0$    &$0.25\pm 0.01$ &$0.50\pm 0.07$
\enddata
\tablecomments{These fits only include quiescent galaxies
(sSFR$<0.3t_H$, where $t_H$ is the age of the universe at each 
redshift (see Appendix); red points in Figure~\ref{fig_smass}). Note that 
there may be systematic offsets between the SDSS and high-redshift sizes 
and/or masses.  Best-fit parameters are defined as
$\log r_e=A+b (\log M/M_\odot-11)$, so that $10^A$ is the typical radius of
a $10^{11}$\,M$_\odot$ galaxy in kpc.}
\end{deluxetable}

\begin{figure*}
\epsscale{1.1}
\plotone{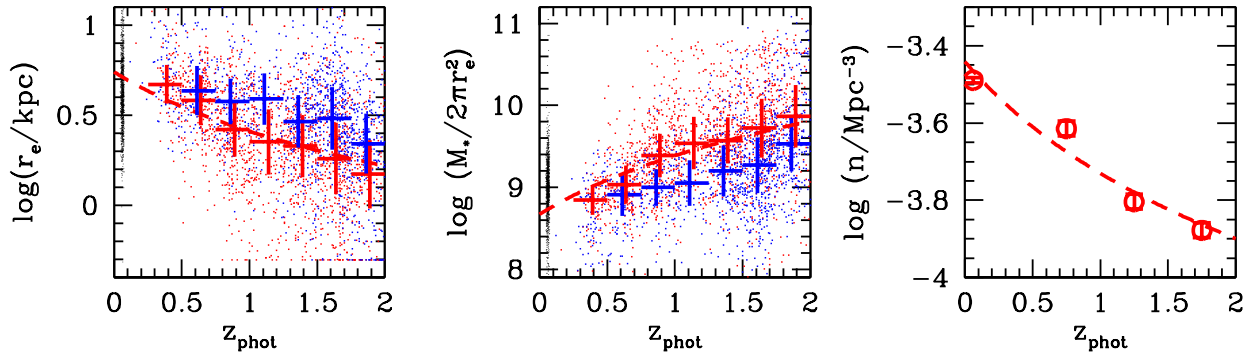}
\caption{{\it Left}: Evolution of $\log(M_\star/M_\odot)>10.8$
quiescent ({\it red points}) and
star-forming ({\it blue points}) galaxy effective radii.  Here
``quiescent galaxies" are defined as those falling on the quiescent
red sequence illustrated in Figure~\ref{fig_z12}.
The large blue and red crosses show the median radii of star-forming
and quiescent galaxies respectively in bins of $\Delta z = 0.25$.  The
dashed line is a power-law (in $1+z$) fit to the small red points.  Small
black points near $z=0.06$ are from the SDSS; note that these points were \emph{not} included in the 
power-law fit.  
{\it Center}: Same as the left panel,
but for galaxy surface densities.  {\it Right}: Evolution of the
massive quiescent galaxy number density with redshift.  Error bars for the
$z=0.5-2$ points do not take cosmic variance or selection uncertainties
into account; only Poisson
errors are included.  Here the $z\sim 0$
point is computed from the SDSS red galaxy mass function of \citet{yang09}.
The observed strong evolution since $z=2$ in each of the three quantities 
(size, surface density, and comoving number density) is well-described with 
a power law in $(1+z)^\alpha$, where $\alpha=-1.09$, $2.37$, and $-1.0$ 
respectively.
\label{fig_sizesd} }
\end{figure*}

The trends shown in Figure~\ref{fig_smass} could in principle be 
affected by certain assumptions we have made.
First, a minimum sSFR of $10^{-11}$ yr$^{-1}$ has been imposed as an 
lower limit to what can be measured with SED fitting; however, varying this
limit only serves to shift the apparent ``quiescence threshold'' in 
this figure, without affecting the overall trends.
Again, the absolute radius, mass, and sSFR values may exhibit some systematic
offsets between the SDSS and UDS samples due to different fitting and 
calculation methods, but the \emph{relative} values within each frame (and 
between the three UDS redshift bins) are generally consistent.  
It is also possible that some degeneracies may exist between the mass and
sSFR estimates used here since both stem from the same template fits
(and hence aren't entirely independent quantities).  To test the robustness
of these results, we reproduced Figure~\ref{fig_smass} using the
SWIRE 24$\mu$m data in this field as a fully independent sSFR diagnostic.
A relation between the sSFR and 
$f_{24\mu m}/f_K$  (with a minor correction using $f_R/f_K$ to account
for low-mass blue galaxies) is presented in \citet[][Appendix]{williams09}. 
Using the $24\mu$m data in place of the best-fit SED sSFR
does not significantly change the results shown in Figure~\ref{fig_smass}
(though it is somewhat noisier, due to the shallowness of the SWIRE 24$\mu$m
data).

\section{The Structural Evolution of Star-forming and Quiescent Galaxies} \label{sec_rfsize}
The strong size and surface density evolution of the ``red and dead'' galaxy
population, and the apparent 
``disappearance'' of the most compact objects, 
presents an intriguing puzzle for observers and theorists
alike.  In order to take full advantage of the unprecedented sample
size of the UDS and provide constraints for present and future models,
we now turn to quantifying these trends.

Figure~\ref{fig_sizesd} shows the sizes and surface densities
of $M_\star>10^{10.8}$ M$_\odot$ galaxies as a function of redshift. 
This mass limit is equivalent (given the difference in IMF) 
to the limits employed by \citet{vdokkum08} and \citet{vdwel09}.  
Quiescent galaxies \citep[selected via the ``quiescent red sequence''
method described in the Appendix and by][]{williams09} are shown in red, and 
star-forming galaxies are shown in blue.  
\citet{vdwel09} note that any proposed models for the size evolution
of quiescent galaxies (in particular those relying on dry mergers) 
are additionally constrained
by the co-moving mass and/or number densities of these
galaxies as a function of redshift.  The
right-hand panel of this figure thus shows the number density evolution
of quiescent galaxies with $\log(M_\star/M_\odot)>10.8$; this was calculated
at $z>0.5$ with galaxy counts in three redshift bins from the UDS, and 
at $z\sim 0$ by 
integrating the SDSS Value Added Galaxy Catalog mass function tabulated
by \citet{yang09}.  The plotted UDS number densities only include Poisson 
uncertainties; cosmic variance and selection effects have not been taken 
into account.  Since the UDS sample is more than 99\% 
complete at these somewhat higher masses, no correction to the
best-fit number densities is necessary.

\begin{deluxetable*}{lcccccc}
\tablecolumns{7}
\tablewidth{400pt}
\tablecaption{Best-fit power law parameters\tablenotemark{1} for the structural evolution of
massive galaxies \label{tab_fit} }
\tablehead{
\colhead{Sample} &
\colhead{$\alpha_r$} &
\colhead{$b_r$} &
\colhead{$\alpha_\Sigma$} &
\colhead{$b_\Sigma$} &
\colhead{$\alpha_n$} &
\colhead{$b_n$}
}
\startdata
ALL &$-0.88\pm 0.06$ &$0.73\pm 0.02$ &$1.92\pm 0.11$ &$7.90\pm 0.04$ &\nodata &\nodata\\
\hline
Quiescent red sequence & $-1.09\pm 0.08$ &$0.74\pm 0.03$ & $2.37\pm 0.14$ &$7.87\pm 0.05$ & $-0.96\pm 0.13$ &$-3.44\pm 0.07$\\
Non-(quiescent RS) &$-1.09\pm 0.08$ &$0.88\pm 0.03$ &$2.44\pm 0.15$ &$7.52\pm 0.06$ &\nodata &\nodata\\
\hline
sSFR$<0.3/t_H$ & $-1.17\pm 0.07$ &$0.78\pm 0.02$& $2.52\pm 0.13$ &$7.80\pm 0.04$ &$-0.73\pm 0.14$ &$-3.45\pm 0.08$\\
sSFR$>0.3/t_H$ & $-0.92\pm 0.09$ &$0.85\pm 0.03$ &$2.14\pm 0.18$ &$7.57\pm 0.07$&\nodata &\nodata\\
\hline
$(U-V)>1.5$ & $-1.00\pm 0.06$ &$0.75\pm 0.02$ & $2.23\pm 0.12$ &$7.83\pm 0.04$ & $-0.47\pm 0.19$ &$-3.45\pm 0.10$\\
$(U-V)<1.5$ & $-1.26\pm 0.19$ &$0.98\pm 0.08$ & $2.79\pm 0.40$ &$7.30\pm 0.16$ & \nodata &\nodata
\enddata
\tablenotetext{1}{Power-law parameters are defined such that
$f(z)= b_f(1+z)^{\alpha_f}$, where $f$ is $r_e$, $\Sigma_\star$, or $n$;
uncertainties are estimated with bootstrap resampling.}
\tablecomments{ 
In all subsamples a mass cut of 
$\log(M_\star/M_\odot)>10.8$ has been imposed.
The SDSS data were only included in the fit for $\alpha_n$; the 
$\alpha_r$ and $\alpha_\Sigma$ fits were based solely on UDS data.  
}
\end{deluxetable*} 

\begin{deluxetable*}{lcccccc}
\tablecolumns{7}
\tablewidth{450pt}
\tablecaption{Size evolution power-law fits in different mass bins
\label{tab_massfit} }
\tablehead{
\colhead{} &
\multicolumn{2}{c}{\underline{All}} &
\multicolumn{2}{c}{\underline{Quiescent}} &
\multicolumn{2}{c}{\underline{Star-forming}} \\
\colhead{$\log M_\star/M_\odot$} &
\colhead{$\alpha_r$} &
\colhead{$b_r$} &
\colhead{$\alpha_r$} &
\colhead{$b_r$} &
\colhead{$\alpha_r$} &
\colhead{$b_r$}
} 
\startdata
$10.6-10.8$ &$-0.51\pm 0.07$ &$0.51\pm 0.02$ &$-0.75\pm 0.10$ &$0.44\pm 0.03$ &$-0.77\pm 0.08$ &$0.70\pm 0.03$\\
$10.8-11.0$ &$-0.81\pm 0.07$ &$0.66\pm 0.03$ &$-1.06\pm 0.11$ &$0.66\pm 0.04$ &$-1.10\pm 0.10$ &$0.86\pm 0.04$\\
$>11.0$     &$-1.09\pm 0.09$ &$0.87\pm 0.03$ &$-1.30\pm 0.10$ &$0.90\pm 0.03$ &$-1.32\pm 0.15$ &$1.03\pm 0.05$
\enddata
\tablecomments{ 
Power-law parameters $\alpha_r$ and $b_r$ are defined such that
$r_e(z)=b_r(1+z)^{\alpha_r}$.  Quiescent and star-forming
subsamples are selected via the ``quiescent red sequence'' method
described in the Appendix.
}
\end{deluxetable*} 

The trends seen in these three quantities
are quite well-fitted with power law functions of the form $Y\sim (1+z)^\alpha$, 
where $\alpha=-1.09$, 2.38, and $-1.0$ for $Y=r_e$, $\Sigma_\star$, and
$n$ respectively (best-fit $\alpha$ and normalization values are also listed in 
Table~\ref{tab_fit} for this and other samples).  These fits were performed 
only to individual galaxies
between $z=0.5$ and $2.0$, excluding the SDSS data and points that are
major outliers in effective radius 
($\log r_e>1.1$ or $\log r_e<-0.25$).  However, the SDSS data point 
was included in the fit to the number density 
evolution.  For comparison, \citet{vdwel09} find that effective radii
of early-type galaxies with \citet{salpeter}
masses $>10^{11}$\,M$_\odot$ (equivalent to our mass limit) are $0.54\pm 0.04$
times smaller at $z=1$ than $z=0$ and $0.3\pm 0.1$ times smaller at $z=2$.
In both size and surface density the evolution they infer is somewhat
slower than what we find; this might be a result of different sample
selection methods.  They also derive a comoving
\emph{mass} density equal to $35\pm 13$\% of the local value at $z=1$ and
$10^{+4}_{-6}$\% at $z=2.4$, while our \emph{number} density evolution is
somewhat shallower than this (although roughly consistent between
$z=0$ and 1), implying significant mass growth in the existing quiescent galaxy
population over these redshifts.

The above fits include galaxies spanning a somewhat wide range of
masses ($\log M/M_\odot>10.8$), but with this large sample it is 
possible to investigate the growth of galaxies as a function of
stellar mass.  Table~\ref{tab_massfit} lists size evolution power-law 
indices ($\alpha_{r}$) and normalizations ($b_r$) for these same
quiescent and star-forming subsamples (as well as the total sample) in
three narrower mass bins.  A strong trend is immediately
apparent: more massive galaxies undergo significantly faster size evolution
with redshift.  For the total sample 
$\alpha_{r}$ steepens from $-0.52$ to $-1.09$ between stellar masses
of $10^{10.6}$ and $>10^{11}$\,$M_\odot$, and from about $-0.8$ to $-1.3$ 
over the same mass interval for both the quiescent
and star-forming subsamples.  A similar trend was also noted by
\citet{franx08}.  Although this could in principle be due to systematic
galaxy mass overestimates at high redshift, such offsets would need
to be quite large and redshift-dependent (e.g., $\sim 0.2-0.3$\,dex
too high at $z\sim 2$, but correct at $z\sim 0.5$).  We therefore conclude
that there is strong evidence for differential size evolution with mass.

As noted in the Appendix, a number of other physically-meaningful 
definitions of ``quiescence'' are possible and may be simpler to define 
within certain
models.  We thus repeated the power law fits to other $\log M/M_\odot>10.8$
quiescent galaxy samples defined through the other two selection methods
described in the Appendix, i.e. based on low best-fit sSFR ($<0.3t_H$) 
and red
rest-frame colors ($U-V_{\rm rest}>1.5$).  For completeness we performed the
same fits to the size and surface density evolution of 
the complementary samples (star-forming and blue), as well
as all massive galaxies in this sample.  Errors on the power-law
slopes were computed using bootstrap resampling.
The results of these fits are 
tabulated, along with the reference ``quiescent red sequence'' sample,
in Table~\ref{tab_fit}.  The sizes and surface densities
of the different samples evolve at similar rates (differing by no more
than $2\sigma$ for any two ``equivalent''  samples, e.g.~low-sSFR
and quiescent red sequence).  The observed structural evolution thus
appears to be more or less independent of the exact technique used to
select quiescent galaxies.  Interestingly, massive galaxies overall exhibit
slower evolution than either of the star-forming or ``dead'' subsamples;  
this is simply a consequence of the increasing fraction of massive quiescent 
galaxies at lower redshifts.

\begin{figure*}
\plotone{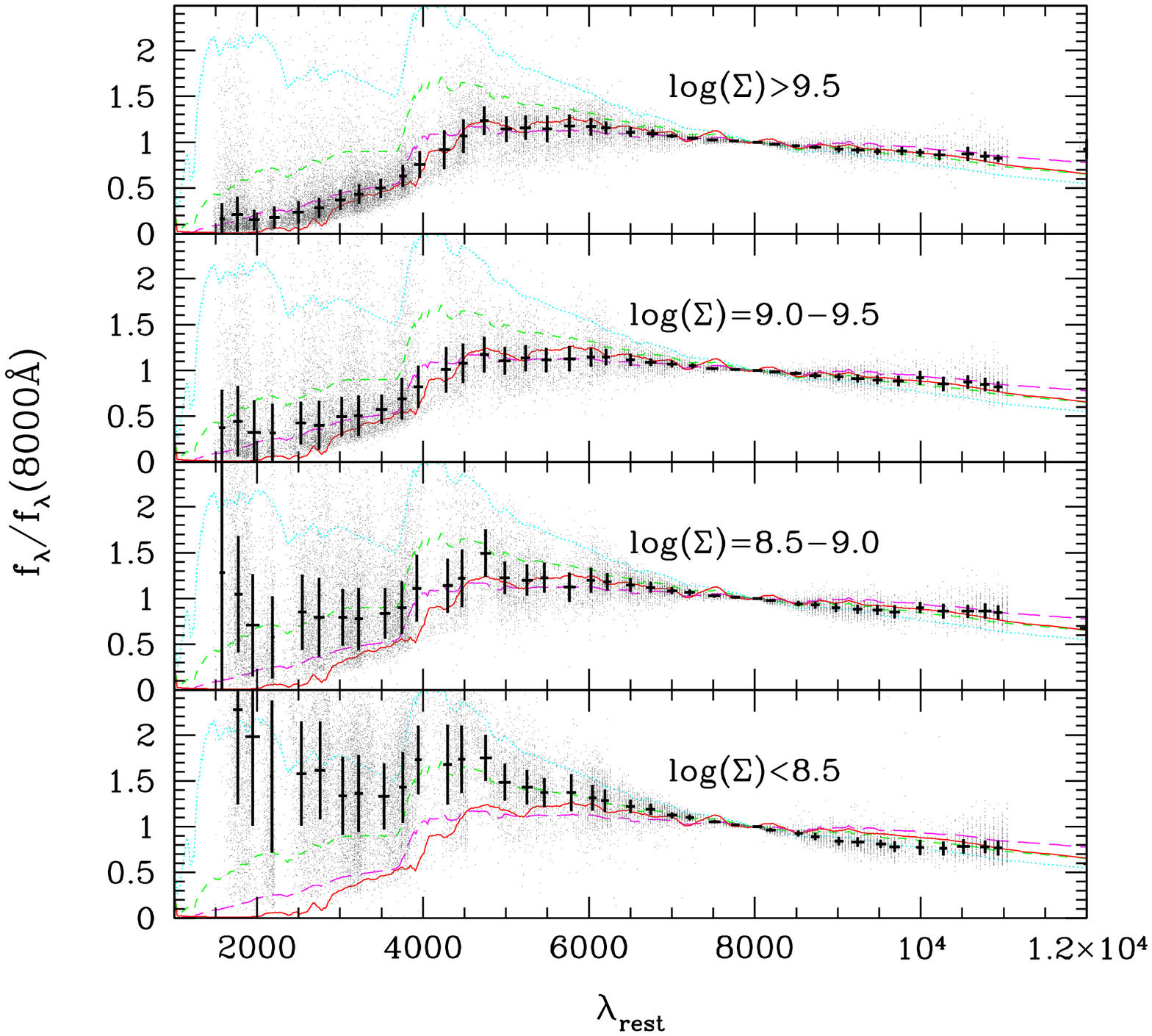}
\caption{Average spectral energy distributions of all $K<22.4$ galaxies
between $z=1-2$ 
in four stellar surface density bins.  Small dots represent broadband flux 
measurements of individual
galaxies, thick black bars denote the median and interquartile dispersion
of the individual points, and overplotted lines show \citet{bc03} 
models of quiescent (1 Gyr age; solid lines) and star-forming galaxies with 
$A_V=1$ 
(dotted), 2 (short dashed), and 3 (long dashed).  The composite SED of the 
highest surface-density galaxies is well-represented by an old or a very
dusty stellar population template.
\label{fig_avgsed}}
\end{figure*}

\section{How universal is the star formation-galaxy structure connection?}
\label{sec_uniformity}
From the preceding sections it is clear that star formation and galaxy
structure are tightly coupled.  While it was already
established that such correlations exist \citep{toft07,franx08}, the 
order-of-magnitude larger galaxy sample provided by the UDS allows
us to not only observe and quantitatively measure said correlations, but 
also to investigate
how uniformly the galaxy population at high redshift is described by them.
In this final section we present some alternate tests
of the connection between structure and star formation, with the specific
goal of qualitatively understanding the degree to which these relations are 
``universal.''

\subsection{Average Spectral Energy Distributions} 
Average SEDs are an invaluable tool for ascertaining
both the underlying spectral shapes of large galaxy samples and the
scatter (or lack thereof) in their SEDs.  To determine whether dense
galaxies are universally quiescent, we separate all $z=1-2$, $K<22.4$ 
UDS galaxies into four {\it surface density} bins and construct average
rest-frame SEDs using their observed, de-redshifted photometric data points.  
Figure~\ref{fig_avgsed} shows these photometric data (normalized
at 8000\AA) in each bin; median and 75\% dispersion values
of the individual galaxy points in bins of rest-frame wavelength 
are overplotted.  To guide the eye, four \citet{bc03} stellar
population models (from bottom to top, one single-burst model with age
1\,Gyr, and three constant 
star-forming with increasing levels of dust obscuration) are also shown
in each panel; note that these are {\it not} fits to the data.

This figure provides a nearly model-independent confirmation of the
result shown in Figure~\ref{fig_smass}: namely, that the densest galaxies
exhibit an average spectral shape that is well-represented by
an evolved stellar population, with remarkably low dispersion.
Interestingly, there appears to be a smooth progression between average
surface density and dust properties: galaxies with the lowest
$\log \Sigma$ are typically blue and have low dust obscuration,
but with increasing surface density the SEDs become dustier.  
It is especially notable that
this does not appear to be caused by the addition of progressively
more quiescent galaxies to the blue galaxy population; at least in the
lowest two $\log\Sigma$ bins, the median SEDs are unambiguously within the
``dusty'' regime.  Thus, it appears that there is not only an anticorrelation
between sSFR and surface density, but also a positive correlation
between {\it dust} and surface density for star-forming galaxies.  However,
since the sSFR is a mass-normalized quantity and the (qualitative) dust
obscuration seen in Figure~\ref{fig_avgsed} is not, we caution that this
last point may be entirely due to more massive galaxies containing more total 
dust.

More importantly, the SED dispersion of the highest-density
galaxies is at least a factor of 2 lower than for less-dense galaxies.
This strongly implies that the highest density galaxies are overwhelmingly
represented by a single model with relatively few outliers; 
at lower surface densities, on the other hand, galaxies show larger
dispersions in their spectra and thus span a range of spectral shapes.   
The comparatively low scatter in the high density galaxies further
reinforces the idea that they are by and large quiescent: if they instead
were predominantly red due to dust obscuration, they would
most likely exhibit large dispersions comparable to the SEDs in the
lower-density bins.  A quiescent galaxy model also provides a somewhat
closer representation of the 4000\AA\ break region than the $A_V=3$ model,
but the models underpredict the observed UV flux in these galaxies, indicating
that there may be some contribution from very dusty starbursts or low
levels of residual star formation in the ``quiescent'' galaxies.

\subsection{Galaxy structure in color-color space}
Up until now we have primarily investigated the star-formation (and SED) 
properties of galaxies as a function of structural parameters.  We now
consider the converse question: what are the structural parameters
of galaxies as a function of their star-formation properties?

\begin{figure}
\plotone{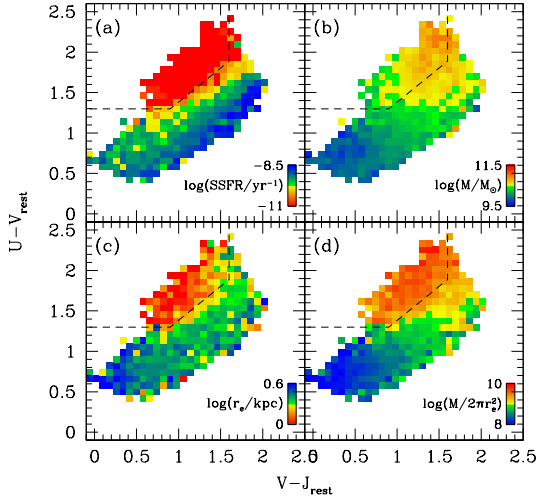}
\caption{
Rest-frame color-color plot for galaxies with 
$1.0<z<2.0$, binned
and color-coded by (a) log specific star-formation rate derived
through SED fitting, (b) log(mass),
(c) log(effective radius), and (d) log surface density.  The dashed
line denotes the quiescent galaxy criterion of \citet{williams09},
where galaxies above the diagonal belong to the ``quiescent red
sequence'' and below the diagonal are star-forming.  
\label{fig_z12}}
\end{figure}

\citet{williams09} describe the use of the rest-frame $U-V$ vs. $V-J$ 
(hereafter $UVJ$) diagram as a powerful diagnostic of both star-formation
activity and dust obscuration.  Figure~\ref{fig_z12} shows this diagram 
for all $z=1-2$ galaxies in the UDS. In short, quiescent galaxies 
fall in a discrete clump in the upper left-hand region of this plot
(above the dashed line), while
star-forming galaxies form a ``dust sequence'' extending from the lower-left
(blue) to upper-right (red) region.  The utility of this plot as a
star-formation diagnostic was confirmed
using stacked $24\mu$m data; a more thorough discussion is
provided in the Appendix.  

In each panel of Figure~\ref{fig_z12} galaxies
have been binned and color-coded by their median (a) sSFR, (b) mass, (c) 
effective radius, and (d) surface density.  In the first panel we confirm 
with our SED-based sSFR estimates what was previously shown by 
\citet{williams09}  with $24\mu$m data
-- the ``quiescent red sequence'' (above the dashed diagonal line) 
is indeed overwhelmingly populated by
quiescent galaxies, while red galaxies that do not lie on this
sequence have significant star formation, and are thus dusty starbursts.
Panels (b) and (c) illustrate how most quiescent galaxies have
large masses and small effective radii, but some systematic variation is 
clearly present (due to the size-mass relation).  However, in the final panel
it is evident that quiescent galaxies almost uniformly exhibit high {\it surface
densities}; there appears to be little variation in surface density with either
mass or size along the quiescent red sequence.  The converse also appears
to hold true: the highest-density galaxies almost entirely fall within
the ``quiescent'' region, confirming the result shown in 
Figure~\ref{fig_avgsed} (although some of the reddest star-forming
galaxies exhibit densities approaching those of quiescent galaxies).

\begin{figure*}  
\plottwo{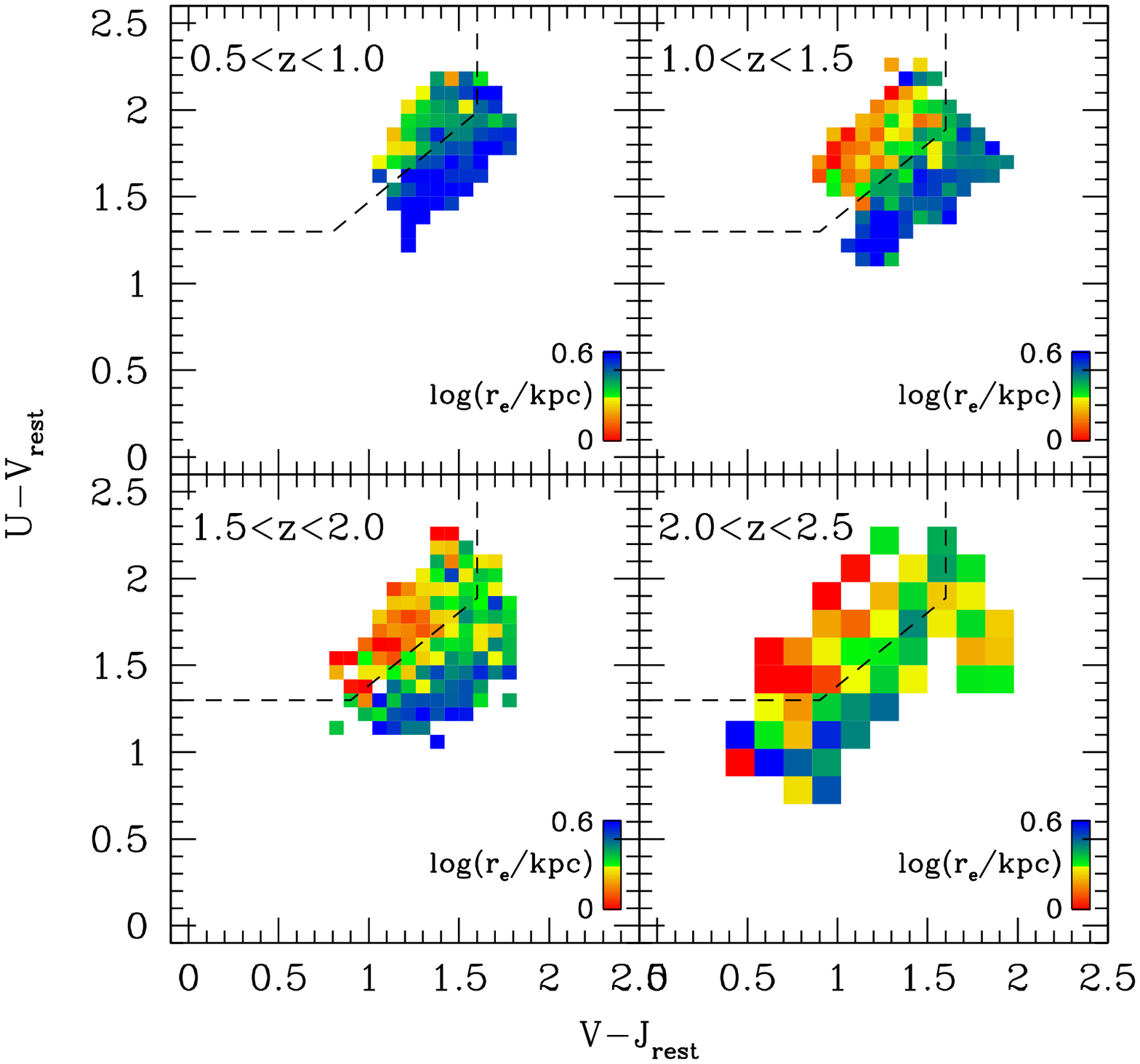}{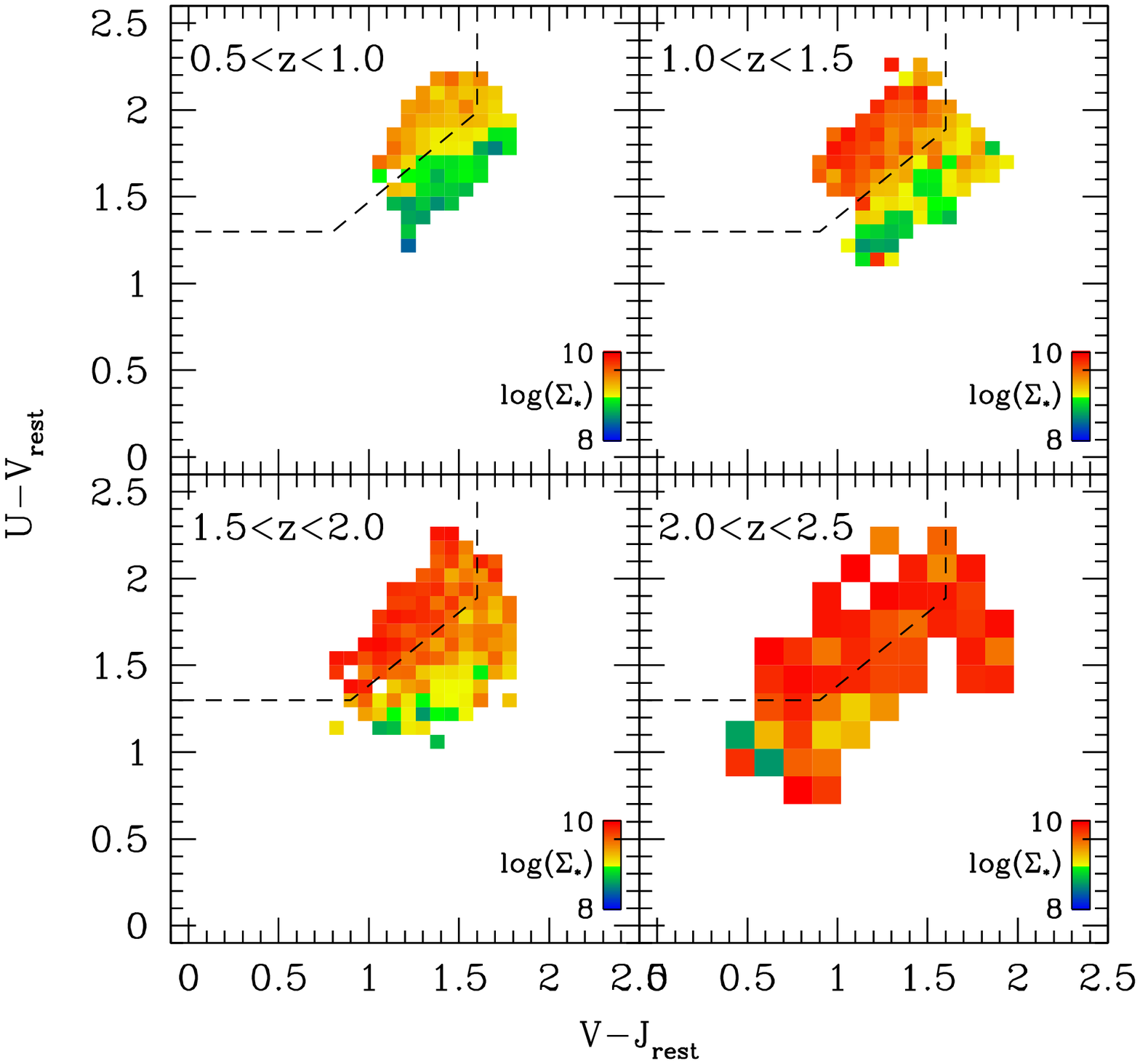}
\vspace{-0.2cm}
\caption{{\it Left panel}: Rest-frame colors of galaxies in four redshift 
bins, color coded by median log effective radius.  Quiescent galaxies
lie above and to the left of the dashed line.
{\it Right panel}: Same as left, but with points color-coded by log surface
density (in $\log M_\odot/$kpc$^2$).  A mass limit of
$\log M_\star>10.6$ has been imposed; thus the $z=2-2.5$ bin is likely
incomplete.  Quiescent galaxies have distinctly
smaller sizes than star-forming galaxies at all redshifts; however, 
at $z>1$ there appears to be a population of dense, star-forming galaxies.
\label{fig_rfcol_sd}}
\end{figure*}

Though useful for illustrating the utility of the $UVJ$ diagram, 
Figure~\ref{fig_z12} spans a fairly wide redshift range and may also contain 
systematic effects due to incompleteness at low masses.  We therefore
re-plot panels (c-d), the size and surface density as a function of
color, in Figure~\ref{fig_rfcol_sd} for four redshift bins. A mass
limit of $\log M_\star>10.6$ (the 75\% red-galaxy mass 
completeness limit at $z=1.5-2$) is also imposed; thus, this figure is
complete only up to $z=2$ and we caution that 
the $z=2-2.5$ bin likely suffers from some incompleteness effects as
well as larger uncertainties in the masses and rest-frame colors.

\vspace{1.5cm}

The color-color plots shown in Figure~\ref{fig_rfcol_sd} present a wealth
of information about the structural evolution of massive galaxies,
and indeed provide a novel view of the results presented
thus far.  First, at all redshifts, quiescent galaxies (above the dashed
line in each plot) exhibit small radii and large surface densities,
though their surface densities are far more uniform than their sizes.
However, while most massive compact galaxies are quiescent, the same
is not always true for massive, high-surface density galaxies: at $z\sim 2$
there are many dense, star-forming galaxies, but the number of such objects
drops dramatically with decreasing redshift, suggesting that galaxies
become quiescent at progressively lower surface densities with time.
Even with this effect taken into account, it remains clear that the quiescent
galaxy population undergoes dramatic structural changes from $z=2$ to 0,
so another mechanism must be at work to bring about this transformation.

\section{Discussion}
\subsection{Comparison to previous work}
By combining size measurements of a very large $K$--selected galaxy sample
with SED fitting to determine SFRs and masses,
we find that quiescent galaxies at all redshifts are far more compact 
than star-forming
galaxies of comparable mass.  Similar results have been found by other
authors at high redshift 
\citep[e.g.,][]{daddi05,toft07,franx08,buitrago08,cimatti08}, but 
with the larger sample size considered here, detailed investigations into 
the correlations between galaxy structure and star formation (and their
evolution with redshift) are now possible.  As shown in 
Figure~\ref{fig_smass}, a well-defined mass-size relation for quiescent 
galaxies exists up to at least $z=2$, but evolves to larger sizes with 
decreasing redshift.  This figure also shows a clear anticorrelation 
between star formation activity and stellar mass surface density (nearly
independent of mass), and an evolving surface density ``threshold''
above which galaxies are predominantly quiescent, confirming
the findings of \citet{franx08}.  

The effective radii of $\log M/M_\odot>10.8$ quiescent galaxies in this 
sample evolve 
as $\sim (1+z)^{-1.1}$ between $0.5<z<2$, with the power-law index varying 
only slightly
depending on the exact definition of quiescence (see Table~\ref{tab_fit}); 
this is in excellent
agreement with the $(1+z)^{-1.22\pm .15}$ ``upsizing'' observed by
\citet{franx08} in a smaller sample from $z=0-3$, and the factor
$4.3\pm 0.7$ size evolution since $z=2.3$ seen by \citet{buitrago08}.  
However, the slope
of the size evolution varies strongly with \emph{mass}, such that the
most massive galaxies show the fastest mass evolution (see 
Table~\ref{tab_massfit}).  \citet{trujillo07} found evidence for similar
mass-dependent evolution of ``disk-like'' galaxies between $0<z<1$; the 
sample presented here confirms that this result extends to higher redshifts
and applies to the entire quiescent galaxy population.

The evolution in the UDS sample also agrees 
well with other previous studies specifically focusing on early-type galaxies;
for example, \citet{vdwel09} find a factor of $\sim 2$ growth in size from 
$z=1$ to the present and a factor of $\sim 3$ since $z=2$, which agrees
with the observed evolution in our ``quiescent red sequence'' 
sample.  On the other hand, we find all galaxies above
$\log M=10.8$ exhibit size evolution with a power-law index of $-0.88\pm 0.06$,
slightly faster than the \citet{franx08} index of $-0.71\pm 0.07$; however,
this may simply be due to the different redshift ranges considered.  The
total massive galaxy sample grows more slowly in size than either 
quiescent or star-forming subsamples of similar mass, 
highlighting the increasing dominance of ``dead'' ellipticals over starbursts 
at lower redshifts.

In the highest redshift bin, $z=1.5-2$, we see a population
of ultra-compact, high-density quiescent galaxies that are no longer 
present in the local universe, or even (for the most part) at $z=0.5-1$.
Studies with higher-resolution data find that
the typical sizes of $M>10^{11}$\,M$_\odot$ quiescent galaxies at $z\sim 2.3$ 
are $r_e\sim 1$\,kpc \citep{vdokkum08}, and at $z>2$ no 
quiescent galaxies with $r_e>2$\,kpc are seen \citep{toft07}.  Although
1\,kpc is smaller than what we can confidently measure for individual
galaxies, this size is consistent with our measurements.
Figure~\ref{fig_smass} does, however suggest the presence of quiescent
galaxies at $1.5<z<2$ with somewhat extended radii ($r_e=2-3$\,kpc).
A few such objects are also seen at similar redshifts (and with
space-based imaging) by \citet{damjanov09}; thus, the difference may simply
be due to mild
evolution of these galaxies from $z\sim 2.3$ to $z=1.5$.  Follow-up
observations of candidate {\it extended} quiescent galaxies at $z\sim 2$
would confirm or rule out their existence, and perhaps shed some light
on the mechanism behind the size evolution.

\subsection{What causes the structural evolution?}
The presence of ultra-compact massive galaxies with extreme surface
densities \citep[and velocity dispersions;][]{vdokkum09} at $z>2$
is puzzling, since the prevalence of such objects is already diminished
substantially by $z\sim 1.5$ \citep{franx08,cenarro09,cappellari09} and they
effectively no longer exist at $z=0$ 
\citep{trujillo09,taylor09b}.  Major and minor mergers, in particular
those involving only gas-poor galaxies (``dry mergers''), likely play a
significant role in both the mass and size evolution of these objects
\citep{khochfar08,hopkins08,feldmann09}.  
However, as suggested by \citet{vdwel09}, other processes are 
also likely to be important, in particular the quenching
of progressively larger star-forming galaxies at lower redshift.  
Indeed, this may account for the seemingly incongruous result that the
quiescent and star-forming samples each evolve faster than the overall galaxy
population.  In other words, if the compact, high-surface-density ``tail''
of the star-forming galaxy population is preferentially quenched at any
given redshift, this will accelerate the apparent evolution of both
subsamples.

The strong size growth of both massive star-forming and
quiescent galaxies by itself suggests that major gas-poor (``dry'') mergers
are not the only process driving massive galaxies' strong structural
evolution.  Galaxies which are undergoing significant star-formation 
must contain substantial amounts of gas, and therefore by definition
cannot undergo dry mergers.  However, the sizes and densities of
massive star-forming galaxies are seen to evolve at a rate similar to the 
quiescent galaxies.  It is therefore possible that at least one of the 
mechanisms acting on compact quiescent galaxies similarly affects
star-forming galaxies.  

Another proposed mechanism is adiabatic expansion, 
whereby mass loss
from evolved stars decreases the potential well depth and therefore
increases effective radii as galaxies age.  As discussed 
by \citet{damjanov09}, however,
such a mechanism does not seem feasible because only a small fraction
of the total mass is expected to be lost in passively-evolving stellar
populations between $z\sim 1.5$ and $z=0$.  The size growth may
also be explained by many minor mergers or the accretion of relatively
low-mass satellites.  \citet{naab09} point out that minor mergers
are more efficient at increasing the radii of the primary galaxy than
major mergers, per unit mass of the secondary galaxies.  This scenario
is also particularly attractive because it would depend only weakly (if at
all) on whether or not the central galaxy is forming stars, and therefore
could explain the size growth of both massive and quiescent galaxies.  
In a forthcoming paper we will investigate the role of major and minor 
mergers on massive galaxy evolution (R.~J.~Williams et al., in preparation).

Although it is yet unclear what causes the observed size and surface density
evolution of massive galaxies, the results presented herein illustrate
how large, ground-based, near-IR surveys can
provide large statistical samples, with sufficiently accurate
size measurements, for comparison with the variety of theoretical models 
now under development.  Upcoming instruments like VISTA and WFC3 will
greatly enhance the volume and quality of available data; however,
the proliferation of complementary observational results from existing 
surveys already present an important challenge to, and constraints
on, these models.

\section{Summary}
By applying accurate size measurements and broadband SED fitting to
the largest photometric sample of massive galaxies from $z=0.5-2$ 
to date, we have investigated in detail the interplay between star formation, 
galaxy structure, and mass, and its evolution with redshift.  Our main 
conclusions are as follows:
\begin{enumerate}
\item{Galaxies with low sSFRs follow a 
well-defined mass-size relation up to at least $z=2$, and this relation
moves to larger sizes at lower redshifts.}
\item{The anticorrelation between stellar mass surface density and star
formation activity is much stronger than the size-sSFR relation; 
this confirms the result of \citet{franx08} that surface density and
star formation are tightly connected at all redshifts.}
\item{Even with the far larger galaxy sample studied here, the densest,
most massive galaxies seen at $z\sim 2$ have essentially
disappeared by $z=0.5-1$.}
\item{The sizes and surface densities of 
massive quiescent and star-forming galaxies evolve smoothly
with time (following simple power-law behavior in $(1+z)$), with more
massive galaxies exhibiting faster evolution.  }
\item{Although galaxies with low surface densities exhibit a wide range
of dust and star-formation properties, at the highest surface densities
their SEDs are well-represented by a single quiescent galaxy model; thus,
these ``dense'' galaxies comprise a relatively homogeneous population.}
\end{enumerate}

\acknowledgments
We thank Natascha F\"orster Schreiber for helpful discussions and the
anonymous referee for suggestions that helped improve the manuscript.  
R.J.W. acknowledges support from the Netherlands Organization for Scientific
Research (NWO) and the Leids Kerkhoven-Bosscha Fonds.  R.F.Q.~is supported
by a NOVA postdoctoral fellowship and S.T. gratefully acknowledges support 
from the Lundbeck Foundation.

\appendix

\section{Defining quiescence}
\begin{figure}
\plotone{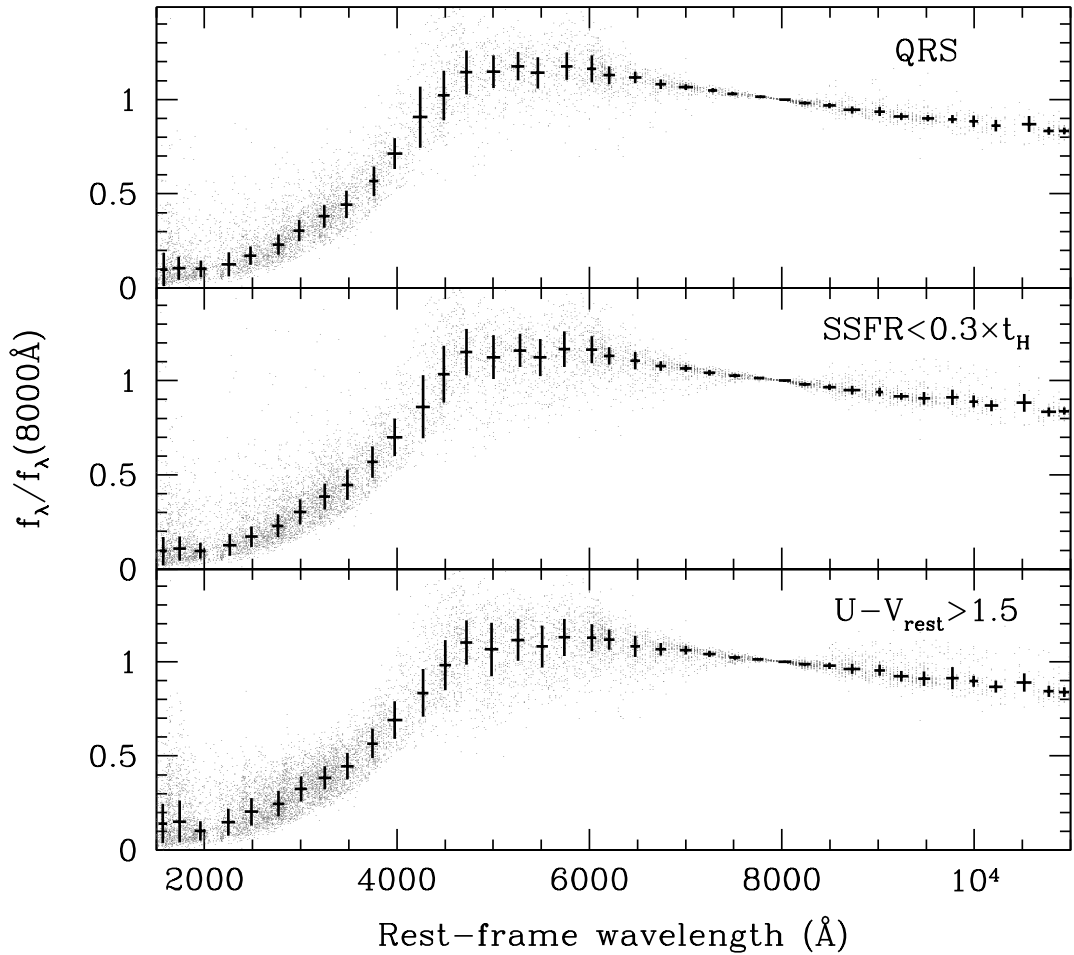}
\caption{
Average spectral energy distributions of $z=1-2$ galaxies
selected via the three methods described in the Appendix: 
the quiescent red sequence in the $UVJ$ diagram 
({\it top panel}); sSFR$<0.3/t_H$
({\it center panel}); and a simple $U-V>1.5$ ``red galaxy'' cut
({\it bottom panel}). Black bars show the median and dispersion of the
individual points in wavelength bins.  
\label{fig_asedred}}
\end{figure}

The primary goal of this work is to investigate the structural properties
and evolution of quiescent galaxies, defined as those falling below some 
sSFR threshold.  In practice there
are a number of ways to define such a sample, particularly when 
only broadband photometric data are available.  In this Appendix
we describe the primary selection technique employed for this analysis
\citep[the $UVJ$ color-color selection described by][]{williams09},
as well as two alternate methods considered in \S\ref{sec_rfsize}.
Average SEDs of the quiescent galaxies selected via these three
techniques are shown in Figure~\ref{fig_asedred}.  Note that these
three average SEDs appear quite similar, suggesting that the three
techniques are effective at selecting quiescent galaxies (though we
caution that very dusty galaxies would not clearly stand out in these
plots; see the models in Figure~\ref{fig_avgsed}).

\subsection{Quiescent Red Sequence}
For the most part, galaxies which have ceased their star formation
activity appear red due to their evolved stellar populations,
and up to $z\sim 1.5$ or higher (depending on the data quality) 
form a well-defined ``red sequence'' 
when plotted in a color-magnitude diagram.  However, at higher redshifts
progressively larger numbers of dusty starbursts are present, and these
often exhibit optical colors which mimic ``red and dead'' galaxies, making
the red sequence methods less reliable (especially with only broadband
photometric data).  As shown by \citet{wuyts07}, incorporating rest-frame
near-IR data into the analysis makes it possible to distinguish between
dead and dusty galaxies.  In particular, quiescent and star-forming 
galaxies occupy distinct regions of the rest-frame $U-V$ vs.~$V-J$ ($UVJ$) 
color space; with the UKIDSS UDS and overlapping data, \citet{williams09}
found that a ``{\it quiescent} red sequence'' is clearly visible (and
distinct from dusty star-forming galaxies)
in the $UVJ$ distribution up to $z=2$, 
and criteria based on these colors provide an effective physical basis
for separating quiescent from star-forming galaxies.

Figure~\ref{fig_z12} (left) shows the $UVJ$ diagram for all galaxies between
$1.0<z<2.0$ in the UDS with $K<22.4$.  The quiescent galaxy selection
criterion defined in \citet{williams09} is shown as a dashed line, where 
quiescent galaxies lie above and to the left of the line and star-forming 
galaxies are below.  For reference, the criterion at $z=1-2$ is:
\begin{equation}
(U-V)>0.88\times (V-J)+0.49
\end{equation}
At lower redshifts the criterion changes slightly, shifting upwards
by 0.1 dex (i.e. the $0.49$ becomes $0.59$).  Additional constraints
of $U-V>1.3$ and $V-J<1.6$ are imposed to prevent too many star-forming
galaxies from scattering into the selection region.

We note that some of the galaxies defined via this method as ``passive'' may 
nonetheless still have emission in the mid-IR from 
dusty star formation or active galactic nuclei (AGNs).  When we compare this technique to deep 
$24\mu$m imaging in the Chandra Deep Field-South, we find that 4 out of 
29 ``quiescent red sequence'' galaxies at $z=2$ have significant
$24\mu$m emission.  Hence this classification is not a guarantee that a 
given galaxy has no star formation.  On average, however, as shown by 
\citet{williams09} for this field, the $24\mu$m emission is very low
for these galaxies.

This empirical ``quiescent red sequence'' method is preferred to, e.g., 
SED fitting because
photometric redshifts (as well as interpolated rest-frame
colors based on these redshifts and observed fluxes) are typically 
the best-constrained parameter in broadband SED fitting.  By providing
a directly empirical criterion, the rest-frame color separation is
also less subject to template-dependent systematics  that may
affect SED-based sSFRs.  We therefore adopt
this color-color cut as our primary method for selecting quiescent galaxy
samples.

\subsection{Evolving sSFR cut}
Since the $UVJ$ technique is relatively new and some theoretical
models may better predict SFRs, we consider
a sample (listed as sSFR$<0.3/t_H$ in Table~\ref{tab_fit}) based
on an evolving cut in galaxy sSFRs.  In this case, quiescent
galaxies are defined as those exhibiting sSFRs
less than 0.3/$t_H(z_i)$, where $t_H(z_i)$ is the age of the universe
for each galaxy at redshift $z_i$, analogous to the quiescent galaxy
definitions employed by \citet{franx08} and \citet{fontana09}.  
The factor 0.3 is somewhat arbitrary,
but is comparable to the sSFRs of red sequence galaxies in our
SDSS subsample.

\subsection{Red galaxies}
Finally, one of the simplest criteria that can be defined is a 
single rest-frame color cut, i.e. assuming that all galaxies redder than a 
certain value are quiescent.  As noted before this method is somewhat
flawed because sufficiently dusty starburst galaxies can mimic the
colors of truly ``red and dead'' galaxies.  Nonetheless, for completeness
we construct a third sample with rest-frame $U-V>1.5$ irrespective
of sSFR.

\end{document}